\pdfoutput=1

\documentclass[11pt,twoside,a4paper,cmspaper,final,collab]{cms-tdr}
\def\svnVersion{344078}\def\cmsCernNoTag{CERN-PH-EP/2015-299}\def\cmsCernDate{\today}\def\cmsMessage{Published in the European Physical Journal C as \href{http://dx.doi.org/10.1140/epjc/s10052-016-4083-z}{\doi{10.1140/epjc/s10052-016-4083-z.}}}

\begin{document}\cmsNoteHeader{SMP-14-017}

\hyphenation{had-ron-i-za-tion}
\hyphenation{cal-or-i-me-ter}
\hyphenation{de-vices}
\RCS$Revision: 334615 $
\RCS$HeadURL: svn+ssh://svn.cern.ch/reps/tdr2/papers/SMP-14-017/trunk/SMP-14-017.tex $
\RCS$Id: SMP-14-017.tex 334615 2016-03-19 11:54:19Z alverson $
\newlength\cmsFigWidth
\ifthenelse{\boolean{cms@external}}{\setlength\cmsFigWidth{0.85\columnwidth}}{\setlength\cmsFigWidth{0.4\textwidth}}
\ifthenelse{\boolean{cms@external}}{\providecommand{\cmsLeft}{top}}{\providecommand{\cmsLeft}{left}}
\ifthenelse{\boolean{cms@external}}{\providecommand{\cmsRight}{bottom}}{\providecommand{\cmsRight}{right}}
\ifthenelse{\boolean{cms@external}}{\providecommand{\breakhere}{-\linebreak[4]}}{\providecommand{\breakhere}}{\relax}

\providecommand{\herwig} {{\textsc{Herwig++}}\xspace}
\providecommand{\nlojet} {{\textsc{NLOJet++}}\xspace}
\providecommand{\fastnlo} {{\textsc{FastNLO}}\xspace}
\providecommand{\fastjet} {{\textsc{FastJet}}\xspace}
\providecommand{\root} {{\textsc{root}}\xspace}
\providecommand{\PYTHIAS} {{\textsc{Pythia6}}\xspace}
\providecommand{\PYTHIAE} {{\textsc{Pythia8}}\xspace}
\providecommand{\RooUnfold} {{\textsc{RooUnfold}}\xspace}
\providecommand{\lumieff} {\ensuremath{\mathcal{L}_\mathrm{int,eff}}\xspace}
\providecommand{\alpsmz}{\ensuremath{\alpha_\mathrm{S}(M_{\PZ})}\xspace}
\providecommand{\alpsq}{\ensuremath{\alpha_S(Q)}\xspace}
\providecommand{\ptpart}{\ensuremath{\pt^{\text{part}}}\xspace}
\providecommand{\rbthm}{\rule[-2ex]{0ex}{5ex}}
\providecommand{\rbtrr}{\rule[-0.8ex]{0ex}{3.2ex}}
\newcommand{\as}{\ensuremath{\alpha_\mathrm{S}}\xspace}
\title{Measurement of the inclusive jet cross section in pp collisions at \texorpdfstring{$\sqrt{s} = 2.76\TeV$}{sqrt(s)=2.76 TeV}}

\author{The CMS Collaboration}

\date{\today}

\abstract{
The double-differential inclusive jet cross section is measured as a function of jet transverse momentum \pt and absolute rapidity $\abs{y}$, using proton-proton collision data collected with the CMS experiment at the LHC, at a center-of-mass energy of $\sqrt{s} = 2.76\TeV$ and corresponding to an integrated luminosity of 5.43\pbinv. Jets are reconstructed within the \pt range of 74 to 592\GeV and the rapidity range $\abs{y}<3.0$. The reconstructed jet spectrum is corrected for detector resolution. The measurements are compared to the theoretical prediction at next-to-leading-order QCD using different sets of parton distribution functions. This inclusive cross section measurement explores a new kinematic region and is consistent with QCD predictions.
}

\hypersetup{%
pdfauthor={CMS Collaboration},%
pdftitle={Measurement of the inclusive jet cross section in pp collisions at sqrt(s) = 2.76 TeV},%
pdfsubject={CMS},%
pdfkeywords={CMS, physics, inclusive jets, QCD}}

\maketitle

\section{Introduction}

{\tolerance=5000
Jets are copiously produced in proton-proton (pp) collisions at the LHC.
In the standard model, the hard-scatter{\breakhere}ing interaction between partons inside
the protons is described by perturbative quantum chromodynamics (QCD).
Particle-level predictions, however, require a nonperturbative (NP)
modeling of hadronization and multiple parton interactions in addition
to the QCD calculation.
The predicted rate and kinematics of jet production are
sensitive to the composition of the proton described by the
parton distribution functions (PDF) and to the strong coupling
constant ($\as$). The evolution of PDFs and $\as$ with
the increase in the magnitude of the four-momentum transfer
is determined by the renormalization group equations of perturbative
QCD~\cite{dglap1,dglap2,dglap3}.
Precision measurements of inclusive
jet production cross sections at different center-of-mass energies
can be used to determine PDFs and $\as$ as well as to
search for deviations in their behavior from QCD predictions~\cite{cmspdf}.
Inclusive jet cross section  measurements have been
performed at the LHC~\cite{7tevref,atlas,7tevref2,aliceref}
and  at other high energy
colliders~\cite{ua1xsec,ua1ratio,ua2xsec,ua2ratio,cdfxsec,cdfratio,d0xsec,d0ratio}.
The measurements (up to 592\GeV) presented here extend the jet transverse momentum
reach of the previous studies.
}

In this study, the inclusive  jet production cross section, $\sigma (\Pp \Pp \rightarrow \text{jet} + \mathrm{X}) $,
is measured as a~function of the jet transverse momentum \pt and absolute rapidity $\abs{y}$.
The analysis is performed with data from $\Pp \Pp$ collisions at $\sqrt{s} = 2.76\TeV$
with the CMS experiment corresponding to an integrated luminosity of  5.43\pbinv.
Originally designed as a reference for heavy ion studies,
this data set also provides an opportunity to close the wide gap in jet measurements
between  the Tevatron at 1.96\TeV and the LHC at 7 and 8\TeV.
When combined with the cross section measurements at other center of mass energies
the present measurement can be used to improve PDF constraints.
The data presented in this paper are collected at low instantaneous luminosity
conditions with, on average, 1.2~primary interactions per triggered event.
The measured cross section is
compared to the prediction from a~next-to-leading-order (NLO) QCD calculation,
performed using the \nlojet (v.4.1.3) generator~\cite{nlojetref1,nlojetref2}
implemented in the \fastnlo (v.2.1.0) framework~\cite{fastnloref3}.
NP~contributions to the cross section
are taken into account in the theoretical prediction; electroweak contributions are
negligible~\cite{ewcorrection}.

\section{The CMS detector}

The central feature of the CMS apparatus is a superconducting solenoid
which provides a magnetic field of 3.8\unit{T}.
Within the solenoid volume are a silicon pixel and strip tracker,
a lead tungstate crystal electromagnetic calorimeter (ECAL), and a brass
and scintillator hadron calorimeter (HCAL), each composed of a barrel and
two endcap sections. Forward calorimetry complements the coverage provided
by the barrel and endcap detectors.
Muons are measured in gas-ionization detectors embedded
in the steel flux-return yoke outside the solenoid.
A more detailed description of the CMS detector, together with a definition
of the coordinate system used and the relevant kinematic variables, can be found
in Ref.~\cite{Chatrchyan:2008zzk}.

\section{Jet reconstruction and event selection}
The particle-flow~(PF) algorithm~\cite{CMS-PAS-PFT-09-001,CMS-PAS-PFT-10-001}
is used to reconstruct and identify
individual particles in an event with optimally combined information
from the various subsystems of the CMS detector.  The particles are identified as:
charged hadrons, neutral hadrons, muons, electrons, and photons.
The PF candidates are combined into jets
using the anti-\kt algorithm~\cite{Cacciari:2008gp} as implemented
in the \fastjet software package~\cite{fastjet}.
A wide reconstruction cone with a radius of 0.7 is used to reduce the sensitivity
to final-state radiation.
Particles identified as charged hadrons are assigned the pion mass,
while neutral hadrons are considered massless and  the four-vector sum
of all reconstructed particles in the jet is calculated.
The measurements of jet energy and momentum
in the CMS detector are affected by a number of experimental factors, such as the limited
coverage of the tracking system and the nonlinear calorimeter response.
The tracking system provides superior  jet reconstruction (\ie,
systematic  uncertainties due to energy calibration and resolution)
in the central region of the detector~($\abs{\eta} < 2.4 $).
To correct for the detector response, the measurements are calibrated using reference processes
with well-understood kinematics~\cite{jesref1}. Jet energy corrections are derived using
simulated events, generated with \PYTHIAS (v.6.4, tune Z2*)~\cite{pythia}
and processed with \GEANTfour~\cite{GEANT4}.
The most recent \PYTHIAS Z2* tune is derived from the Z1 tune~\cite{uez1},
which uses the CTEQ5L parton distribution set, whereas Z2* adopts CTEQ6L~\cite{cteq}.
The corrections are verified in data using $\gamma {+}  \text{jet}$ and $\PZ {+} \text{jet}$ processes,
and additional corrections are applied to compensate for any mismatch between
simulation and data.
The correction factors depend on jet \pt and  $\eta$,
and typically range between 1.02 and 1.10, while the jet energy resolution amounts
to 15\% at a jet \pt of 10\GeV, 8\% at 100\GeV, and 4\% at 1\TeV.

The events are selected by a set of single-jet triggers with jet \pt thresholds
of 40, 60, and 80\GeV with the first two triggers being prescaled.
In Table \ref{triggertable}, the effective integrated luminosity collected
with each trigger and the corresponding jet \pt range is presented.
The triggers are
selected to ensure 99\% efficiency for the events in the corresponding
\pt range of the analysis.

\begin{table}[h]
\topcaption{Effective integrated luminosities   and jet \pt ranges for triggers used
in this study.}
\centering
\begin{tabular}{cccc}
\hline
\ifthenelse{\boolean{cms@external}}{
Nominal trigger &   \lumieff & \pt range \\
 threshold ($\GeV$)   &   ($\pbinv$)  & (\GeVns) \\
 }{
 Nominal trigger threshold (\GeVns) &   \lumieff ($\pbinv$)   & \pt range ($\GeV$) \\
 }
\hline
  40   & 0.59                                     & \phantom{0}74--97\phantom{0}   \\
  60   & 3.48                                    & \phantom{0}97--133   \\
  80   & 5.43                                    & 133--592  \\
\hline
\end{tabular}
\label{triggertable}
\end{table}

Events with $\ETm/\Sigma \ET< 0.3$ are selected, consistent with the properties of
QCD multijet events, thereby removing any spurious jet-like features originating from
isolated noise patterns in certain HCAL and ECAL regions. The quantities \ETm and $\Sigma \ET$
are calculated as the negative vector sum of transverse energy and the scalar sum
of transverse energy, respectively, of all  PF candidates in the event.
The selected events are required to have at least one well-reconstructed
primary vertex. Each jet should contain more than one PF candidate.
The fraction of jet energy carried by charged leptons ($\Pe$, $\mu$) should be
less than 90\%. In addition, jets reconstructed
within the acceptance range of the tracking system ($\abs{\eta}<2.4$) must
contain at least one charged particle. The electromagnetic energy
fraction of such jets is required to be less than 99\%, while
the neutral-hadron and the photon energy fractions are required
to be less than 90\%. The jet selection
efficiency is estimated to be 99\% or higher for all \pt and
rapidity ranges used in this study.

\section{Cross section measurement}

The double-differential jet cross section is calculated as
\begin{align}
\frac{\rd^2\sigma_{}}{\rd\pt \rd y} =  \frac{1}{\epsilon\lumieff}\frac{N}{\Delta \pt~(2\,\Delta\abs{y})},
\end{align}
where  $\lumieff$ is the effective
integrated luminosity corrected for trigger prescales,
$\epsilon$ is the overall reconstruction efficiency including
the trigger and  jet selection efficiencies,
$\Delta \pt$ and $\Delta \abs{y}$ are the sizes of a particular
jet \pt and rapidity bin, and $N$ is the number of jets
in that bin. Six uniform bins in $\abs{y}$  are used  between 0.0 and 3.0.
The jet \pt values range from 74 to 592\GeV, with bin sizes increasing
in proportion to the \pt resolution.

In order to facilitate the comparison of measurements with
theoretical predictions, the jet \pt spectrum is corrected for
detector effects.
Since the \pt spectrum is steeply falling, the number of jets migrating
out of a bin into the higher adjacent bin significantly exceeds the number
of jets migrating to the lower adjacent bin.
The unfolding procedure compensates for this effect and recovers
the particle-level spectrum from the observed spectrum.
The detector response function is determined using
multijet events  simulated with the \PYTHIAS~(v.6.4,  tune Z2)~\cite{pythia,uez2}
event generator. A detailed  detector simulation is carried out using
the \GEANTfour software to model the particle
interactions in the detector material.

The detector is characterized by a response function
that represents the probability density to
reconstruct a jet with transverse momentum $\pt^{\text{det}}$
when the particle-level jet transverse  momentum is $\pt^{\text{part}}$.
The response function is initially derived by calculating jet resolution in Monte Carlo (MC) simulation
for every \pt and $\abs{y}$ bin. Jet resolution in data is found to be worse than
in simulations~\cite{jesref1}. The response function is corrected
for this defect by degrading the resolution by factors $c^{\text{data/MC}}$ that vary with $\abs{y}$
as listed in Table~\ref{resfcntable}.

\begin{table}[h]
\renewcommand{\arraystretch}{1.1}
\center
\topcaption{
The factors used to scale jet resolution determined in simulations
to match the resolution observed in data.
}
 \begin{tabular}{cc}
\hline
 $\abs{y}$      & $c^{\text{data/MC}}$     \\
\hline
0.0--0.5 & 1.079 $\pm$ 0.026 \\
0.5--1.0 & 1.099 $\pm$ 0.028 \\
1.0--1.5 & 1.121 $\pm$ 0.029 \\
1.5--2.0 & 1.208 $\pm$ 0.046 \\
2.0--2.5 & 1.254 $\pm$ 0.062 \\
2.5--3.0 & 1.395 $\pm$ 0.063 \\
\hline
\end{tabular}
\label{resfcntable}
\end{table}
The response matrix is constructed by convolving the response function
with the \ptpart spectrum predicted by NLO QCD calculations and the CT10
PDF set~\cite{ct10ref}.
(Results with other PDF sets are discussed in Sec.~\ref{sec:unc}.)
The response function is represented by a kernel density estimation (KDE)
technique that accurately models the tails of the distribution.
The theoretical \ptpart spectrum is fitted with an exponential of a continuously
differentiable function (Akima spline)~\cite{splineref}. This spline function is sampled many
times and convolved with the KDE response function to obtain the response
matrix.  The D'Agostini iterative unfolding method~\cite{agostiniref1} is used, as implemented
in the \RooUnfold software package~\cite{roounfoldref}.  The unfolding procedure is
regularized by early termination of iterations; four iterations are performed
in each rapidity bin.

\section{Theoretical predictions}
\label{sec:theory}

The theoretical predictions are derived at NLO using QCD calculations with
\nlojet~\cite{nlojetref1,nlojetref2}, and corrected for
the NP contributions from hadronization and multiple parton interactions.
Electroweak corrections  are negligible at 2.76\TeV according to the studies  performed
in Ref.~\cite{ewcorrection}. The factorization and renormalization scales are set to the
jet \pt ($\mu_{\mathrm{F}} = \mu_{\mathrm{R}} = \pt$).
The theoretical predictions of the inclusive jet cross section
are derived using five recent PDF sets
at NLO, as listed in  Table~\ref{tab:pdfsets},
with the central values of \alpsmz for each PDF set.
Most are determined in a variable-flavor number scheme,
except for the ABM11 PDF set, which employs a fixed-flavor number scheme with
the number of active flavors ($N_\mathrm{f}$) set to 5 or 6.
The details related to determination of the
PDFs are described in the corresponding references.

\begin{table*}[h]
\center
\caption{The PDF sets used for deriving cross section predictions are given
with the number of active flavors ($N_\mathrm{f}$), the values and ranges of \alpsmz
used for the fits, and corresponding references.
}
\begin{tabular}{ll@{\hspace{10mm}}l@{\hspace{10mm}}l@{\hspace{10mm}}l}
\hline
Base set    & $N_\mathrm{f}$  & \alpsmz & \alpsmz range  & Reference           \\
\hline
CT10        & ${\leq}5$   & 0.118        & 0.112--0.127      &  \cite{ct10ref}     \\
MMHT14      & ${\leq}5$   & 0.120        & 0.108--0.128      &  \cite{mmhtref}     \\
NNPDF3.0    & ${\leq}6$   & 0.118        & 0.115--0.121      &  \cite{nnpdf30ref}  \\
HERAPDF1.5  & ${\leq}5$   & 0.1176       & 0.114--0.122      &  \cite{heraref}     \\
ABM11       & \phantom{$\leq$}5         & 0.118        & 0.110--0.130      &  \cite{abm11ref}    \\
\hline
\end{tabular}
\label{tab:pdfsets}
\end{table*}

The NP effects include hadronization of parton cascades leading
to the formation of color neutral jets and multiple
interactions of spectator partons within the colliding
protons that can result in the appearance of additional
jets. The corrections are derived using two event generators
with different models for parton cascades and hadronization:
\PYTHIAS~(v.6.4, tune Z2)~\cite{pythia,uez2} and
\herwig~(v.2.5.0, tune UE\_EE\_3C)~\cite{herwigref,herwigueref}.
In \PYTHIAS, the hadronization is simulated with the
Lund string fragmentation model~\cite{stringfragmentation}
while \herwig employs the cluster fragmentation model~\cite{clusterfragmentation}.
The  \pt- and $\abs{y}$-dependent correction factors for the NP effects, $C_{\mathrm{NP}}$,
are derived from simulation as a ratio of differential jet cross sections
with hadronization and multiple parton interactions turned on and off.
The final correction factors are obtained by averaging \PYTHIAS and \herwig predictions.
The theoretical cross section is then calculated as
$\sigma_{\mathrm{theory}} = \sigma_{\mathrm{NLO}} \, C_{\mathrm{NP}}$.
The $C_{\mathrm{NP}}$ factors vary between 1.02 and 1.10 in the \pt and rapidity range of this analysis.

\section{Systematic uncertainties}
\label{sec:unc}
The major experimental uncertainties in this analysis come from
 imperfect measurement of jet energy, limited precision in simulating
jet energy resolution, and imprecise knowledge of integrated luminosity.
The first source affects the jet spectrum observed in data, while the second
modifies the detector response matrix used in the unfolding
procedure. The third source, measured integrated luminosity, contributes
an overall cross section uncertainty of 3.7\%~\cite{lumref}.
The uncertainty associated with the jet energy determination consists
of several independent contributions identified in the process of deriving
the jet energy corrections. These contributions are described in
detail in Ref.~\cite{jesref1}. The corresponding cross section
uncertainty is 5--22\% for the low-rapidity bins ($\abs{y}<2.5$),
increasing to 78\% in the highest rapidity bin ($2.5\leq\abs{y}<3.0$).
The jet energy resolution uncertainty is estimated using the
uncertainties in the $c^{\text{data/MC}}$ scaling factors presented
in Table~\ref{resfcntable}. For the rapidity region $\abs{y}<2.5$,
the corresponding cross section uncertainty is 2--3\%,
increasing to 22\% for the most forward rapidity bin.
The higher uncertainty at forward rapidities is caused by
the significant increase in the  jet energy and resolution uncertainties,
and the more  steeply falling \pt spectrum in comparison with the central rapidity region.

The energy offset due to additional interactions in the same bunch crossing (pileup)
is small. For the lowest \pt jets considered (74\GeV) the pileup
contributes an average of only 0.3\% of the energy.
This fraction decreases with increasing \pt.
Consequently, pileup corrections are not required
and the associated uncertainties are negligible.
An uncertainty arising from the potential mismodeling of trigger
and jet selection requirements is found to be  1\%.
The unfolding uncertainty due to the initial theoretical model
is calculated by testing various models and finding the effect is negligible.
The sum in quadrature of all experimental systematic uncertainties
in the cross section is, on average, 6\% at low rapidities ($\abs{y}<2.0$)
and varies from  10\% to 80\% at higher rapidities ($2.0\leq\abs{y}<3.0$),
across the corresponding~\pt ranges.

The uncertainty in the theoretical cross section prediction
is estimated from the PDF uncertainties, the choice
for the  factorization and renormalization
scales ($\mu_\mathrm{F}$ and $\mu_\mathrm{R}$), and the variation in the
modeling of NP corrections. The PDF uncertainty, for all PDF sets except NNPDF3.0,
is calculated as the change in the cross section caused by varying
decorrelated PDF parameters. The relevant PDF eigenvectors are provided in the PDF sets
along with the central values. The uncertainty due to each parameter is
determined at 68\% confidence level (CL), and the resulting asymmetric
uncertainties are combined in quadrature.
In the case of NNPDF3.0, the PDF set contains an ensemble of replicas
corresponding to one standard deviation in the PDF.
The PDF uncertainty is calculated by evaluating the standard
deviation in the cross section derived by using different replicas.
The uncertainty due to the variation of the value of \alpsmz in the PDF sets
is found to be much smaller than other uncertainties (${<}1\%$)
and is not included.
The scale uncertainty is determined by varying the factorization and renormalization
scales with respect to the nominal value ($\mu =$ jet \pt)
using the following combinations of ($\mu_\mathrm{F}/\mu,~\mu_\mathrm{R}/\mu$) ratios:
(0.5, 0.5), (1, 0.5), (0.5, 1), (1, 2), (2, 1), and (2, 2).
The largest deviation from the nominal cross section, found
separately in each \pt and $\abs{y}$ bin, is taken to represent
the scale uncertainty. The scale uncertainty is asymmetric and its distribution is
skewed towards lower cross sections.
The largest deviation from the average value of
the $C_{\mathrm{NP}}$ correction factors,
which are obtained with the \PYTHIAS and \herwig generators as discussed in
Section~\ref{sec:theory}, is used as the measure of the NP modeling uncertainty.
It contributes a 2--5.6\% uncertainty in the cross section prediction.
The  uncertainties in the theoretical predictions differ for each PDF set considered, and
typically vary in the 10--20\% range over most of the kinematic region.

\section{Results}
\label{sec:results}
\begin{figure*}[h]
\center
\includegraphics[scale=0.7]{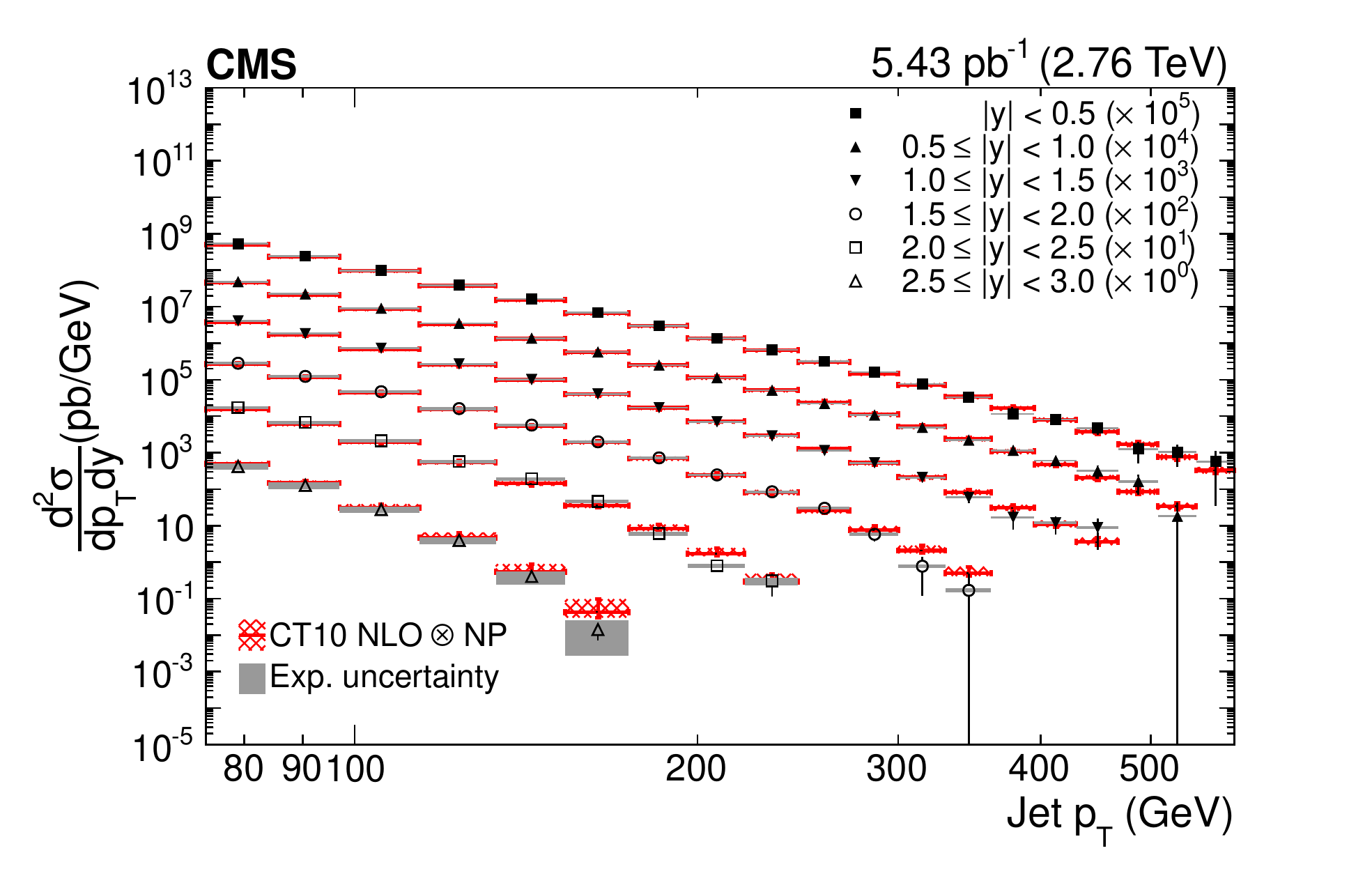}
\caption{The inclusive jet production cross section, measured at $\sqrt{s}= 2.76\TeV$,
shown as a function of jet \pt in six $\abs{y}$ bins, as indicated by different symbols.
The statistical (systematic) experimental uncertainties are indicated by vertical error
bars (filled bands). The measurements are compared to the NLO QCD prediction using
CT10 PDF set. The theoretical uncertainties are represented by hatched bands.
}
\label{combct10}
\end{figure*}

\begin{figure*}[]
\center
\includegraphics[scale=0.35]{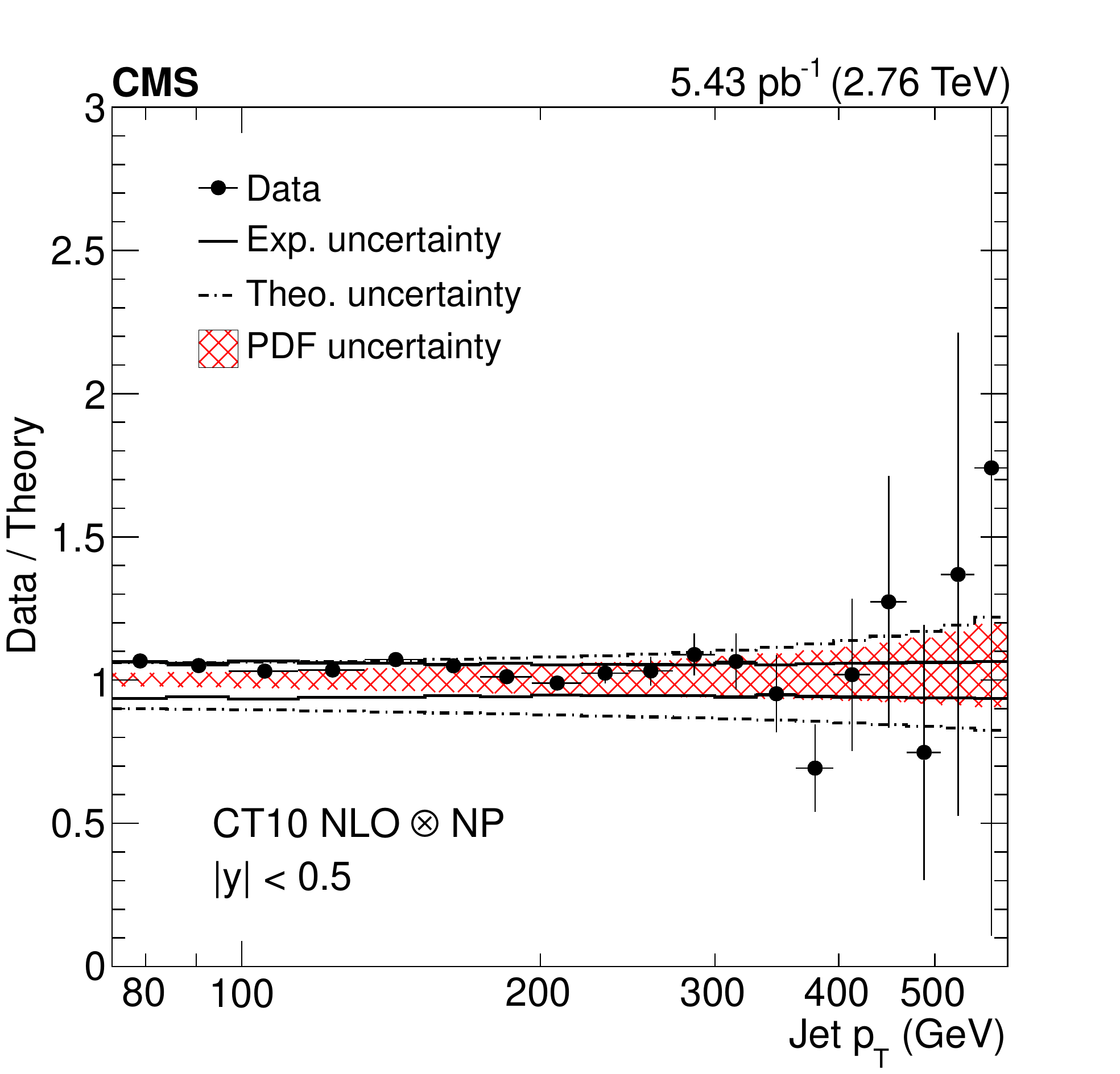}
\includegraphics[scale=0.35]{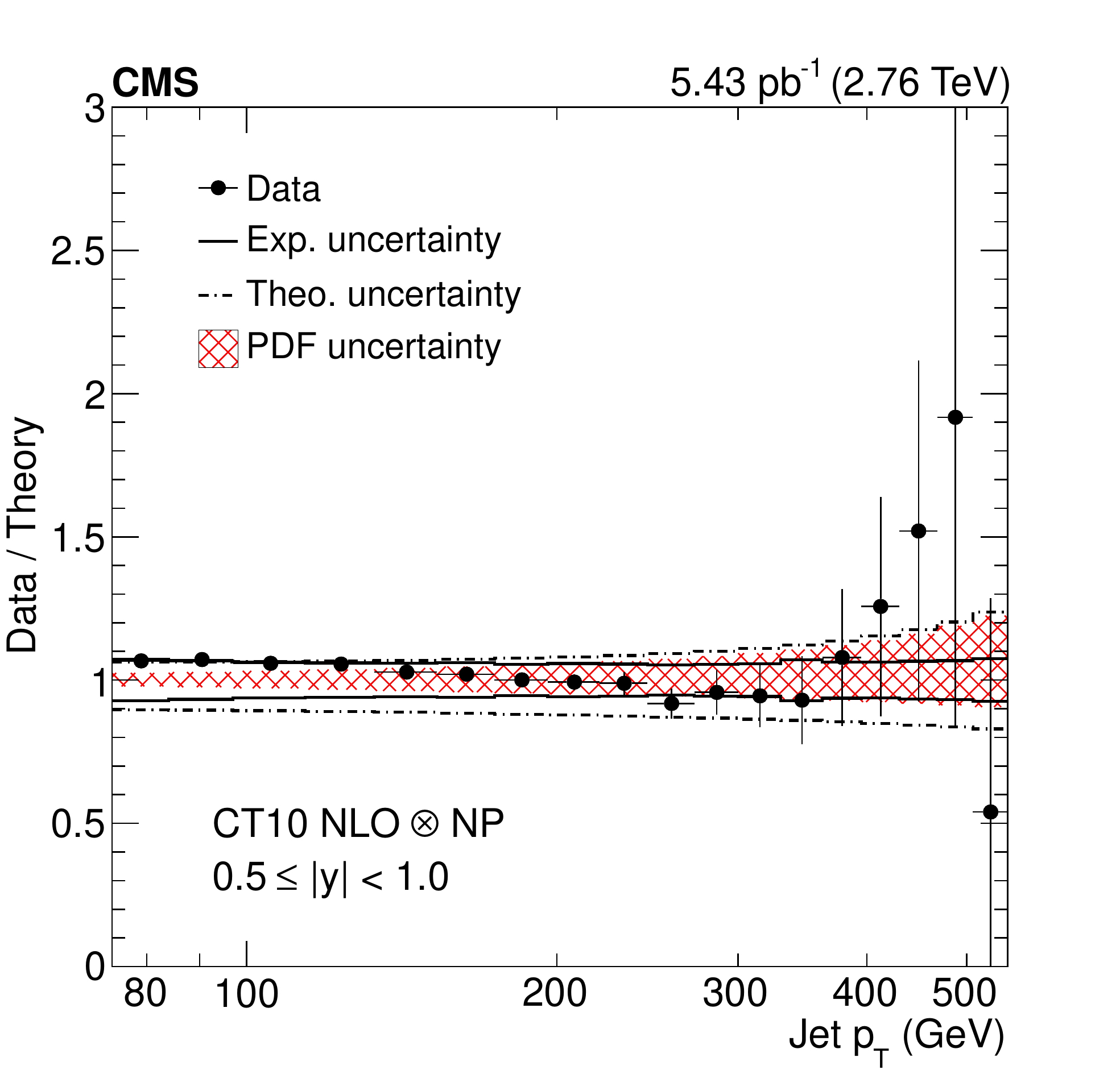} \\
\includegraphics[scale=0.35]{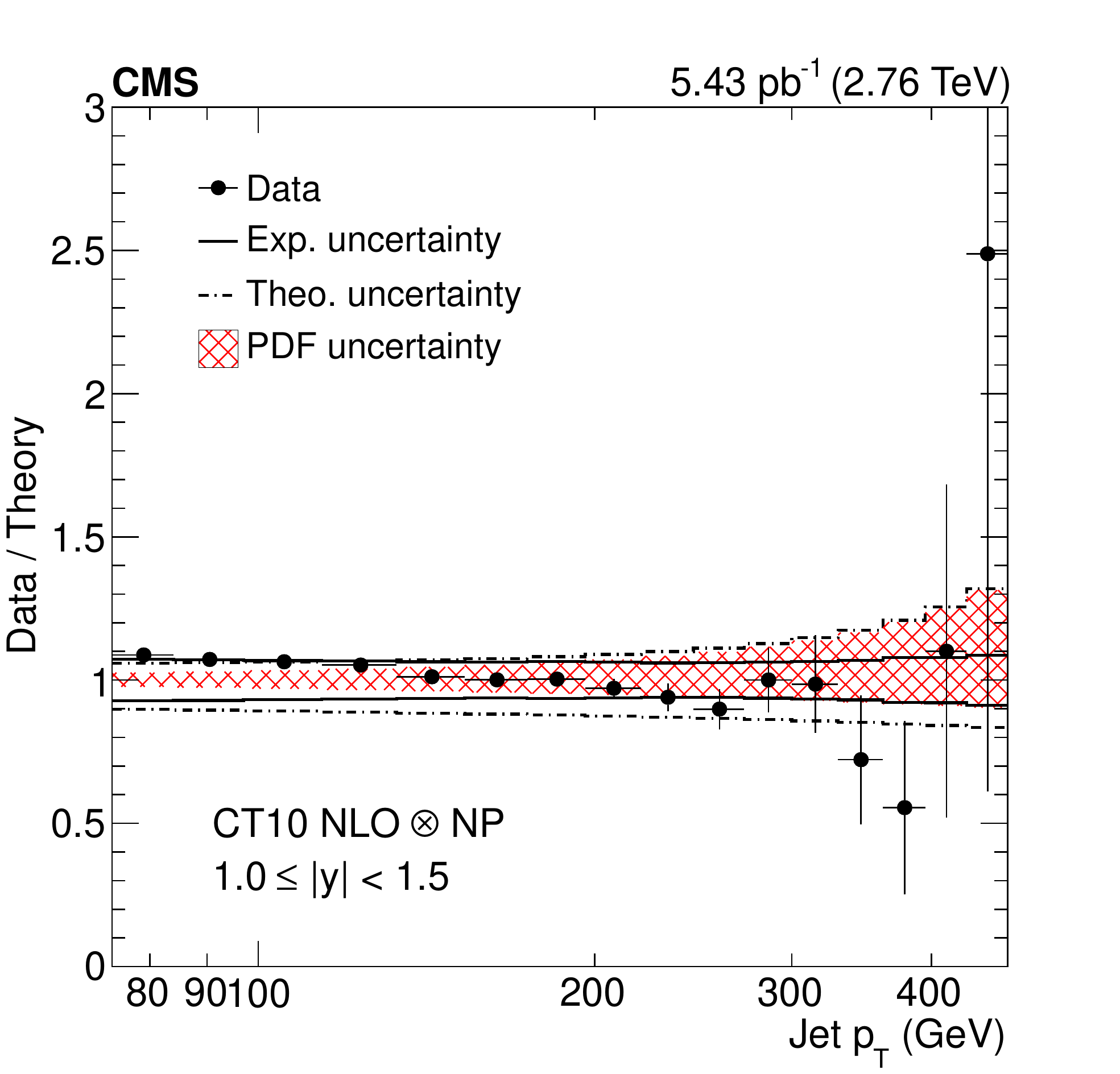}
\includegraphics[scale=0.35]{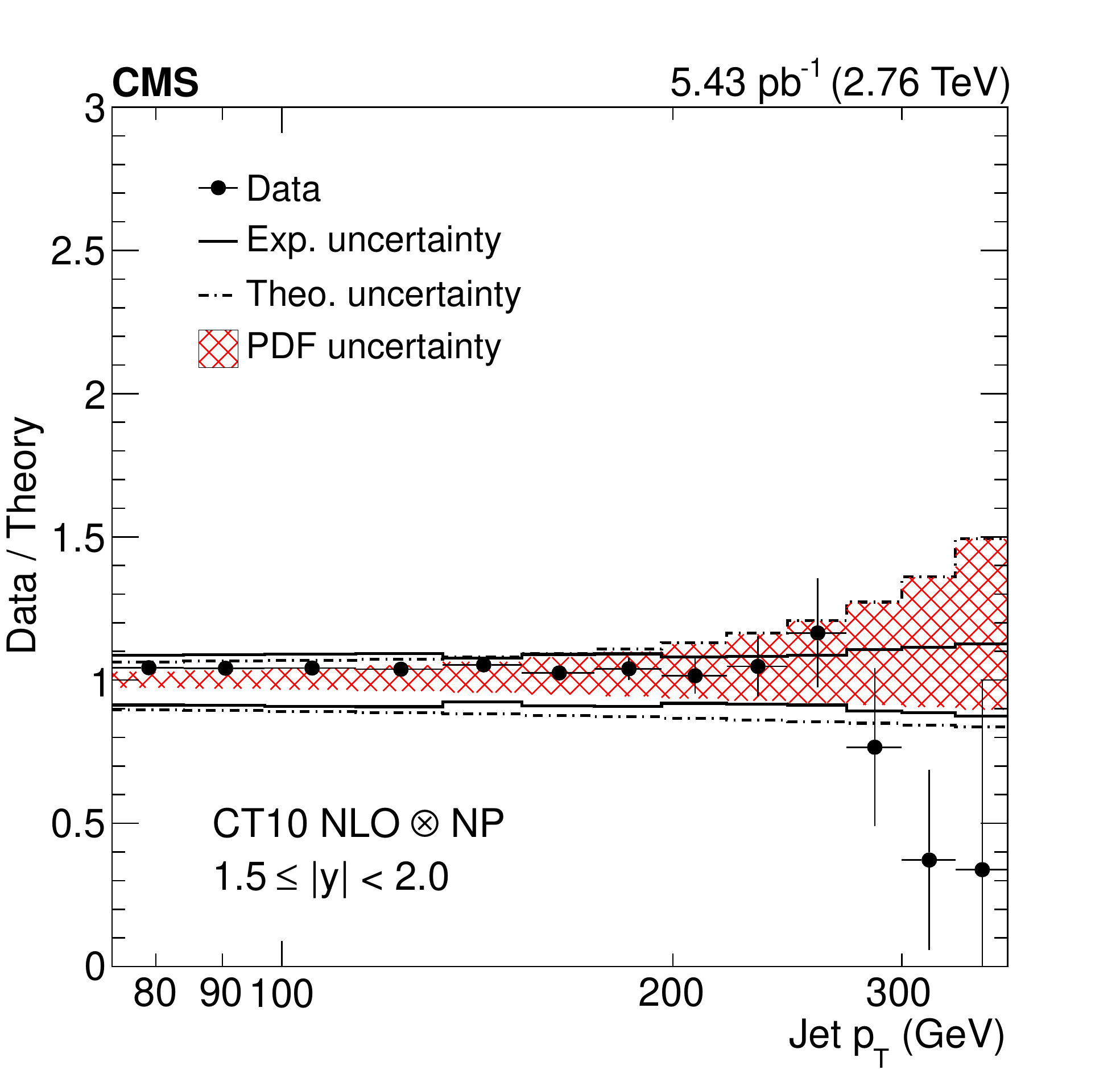} \\
\includegraphics[scale=0.35]{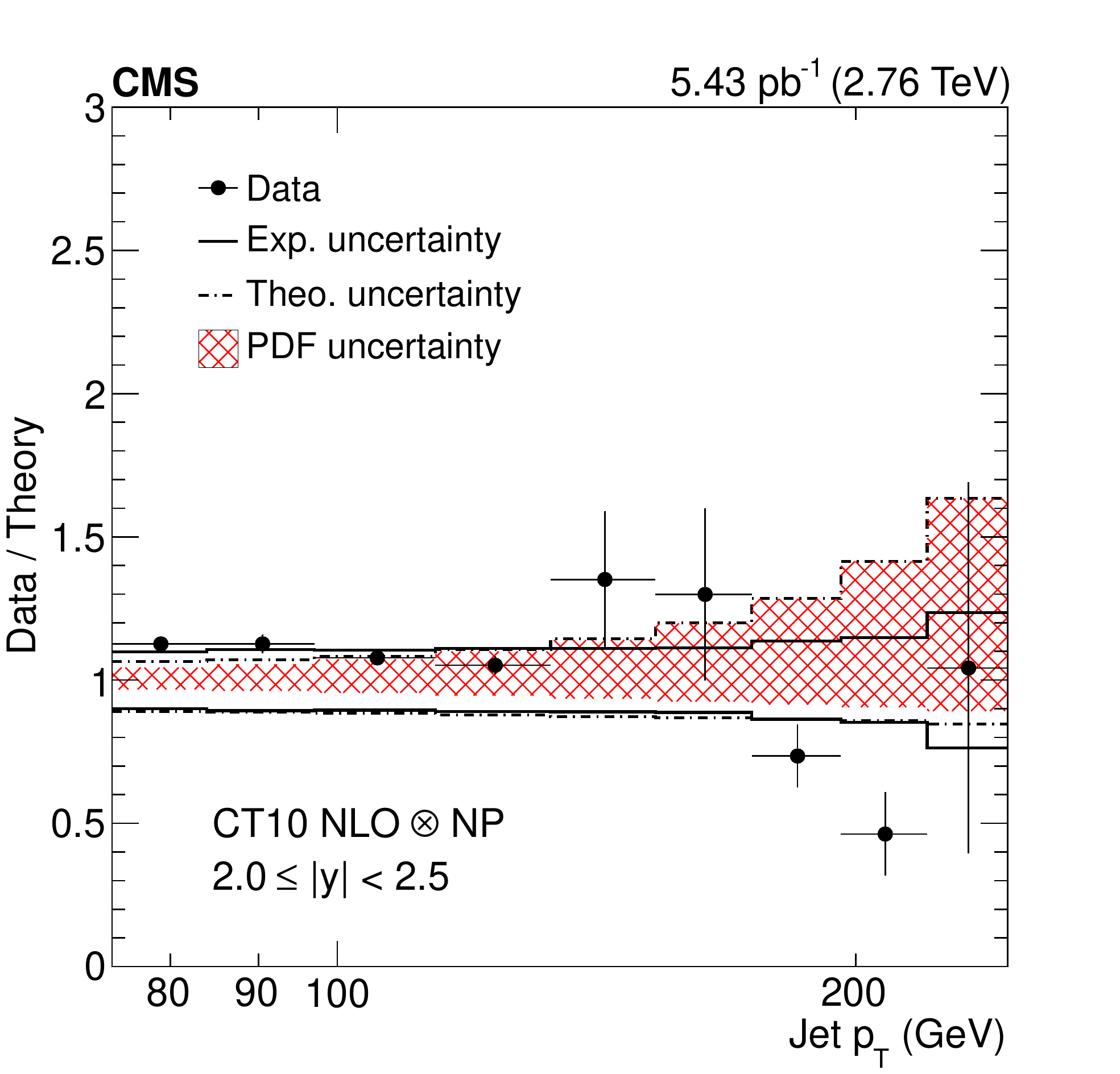}
\includegraphics[scale=0.35]{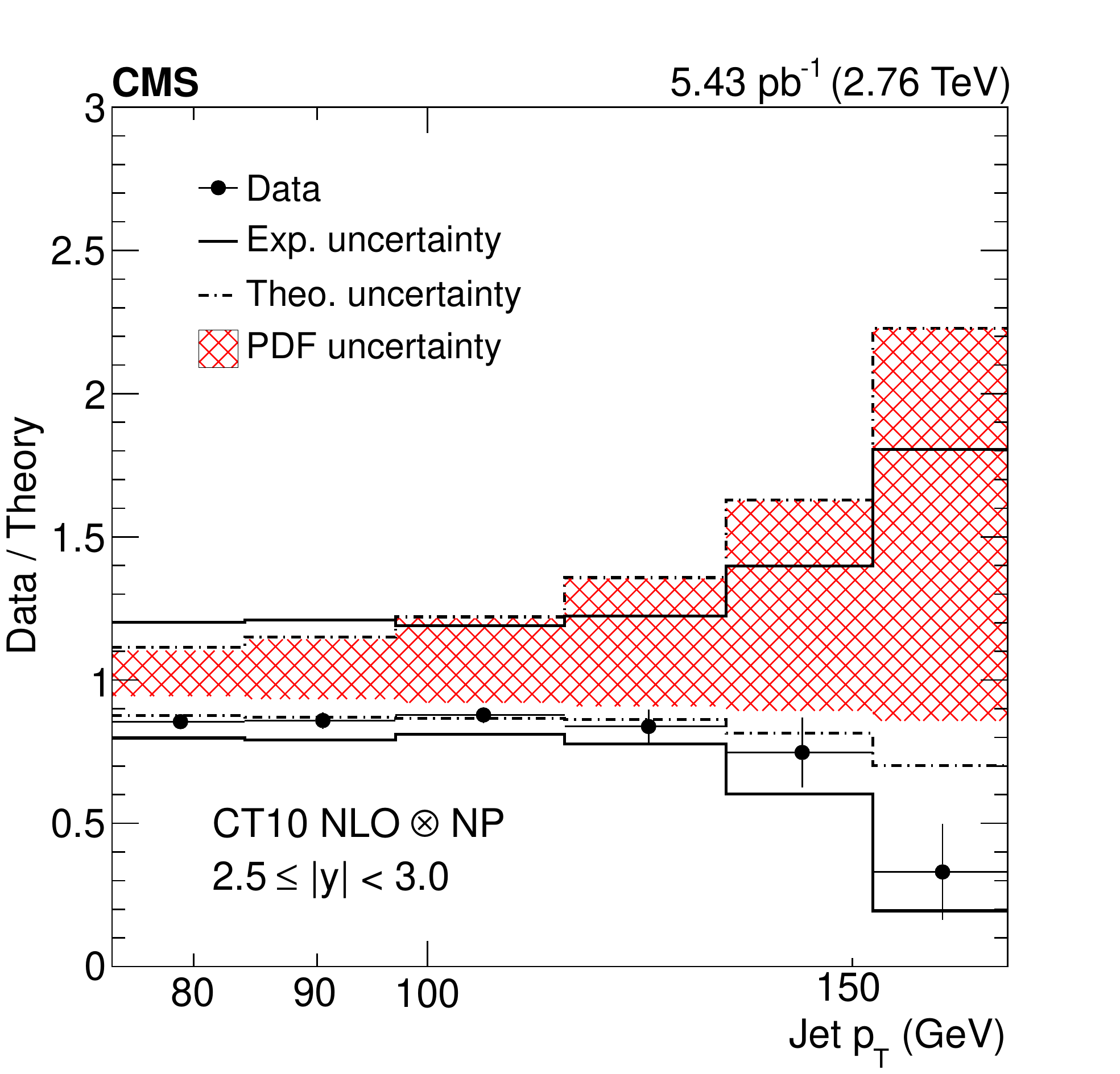}
\caption{The ratio of the measured inclusive jet production cross section
(closed symbols)  at $\sqrt{s}= 2.76\TeV$ to the theoretical
prediction using the CT10 PDF set is shown as a function of jet \pt in each
measured $\abs{y}$ range with the statistical (vertical error bars) and
systematic (solid lines) experimental uncertainties. The total theoretical
uncertainties are shown by the dash-dotted lines with the contribution
from PDF uncertainties (hatched band).
}
\label{datathyct10}
\end{figure*}

\begin{figure*}[]
\center
\includegraphics[scale=0.35]{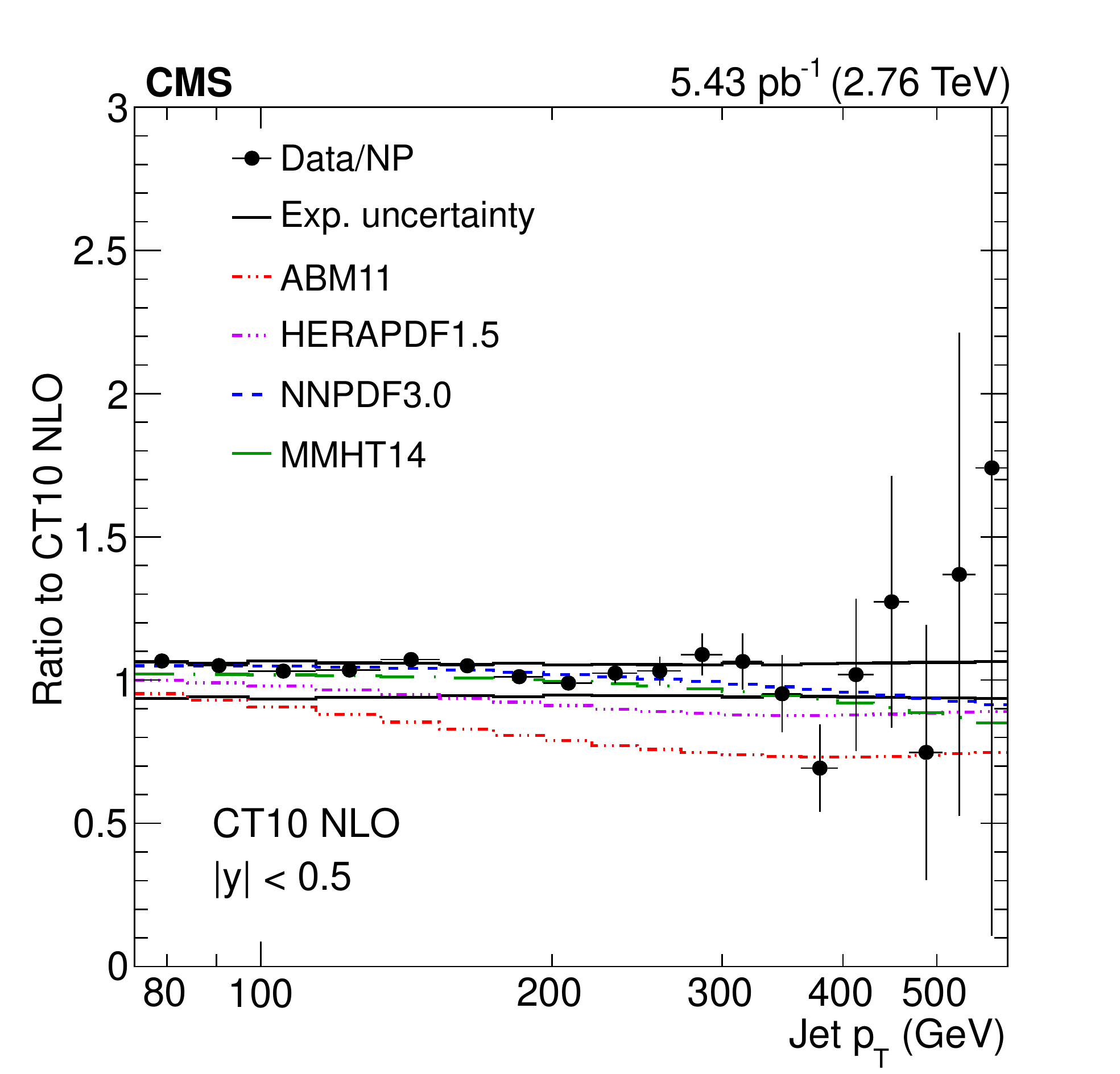}
\includegraphics[scale=0.35]{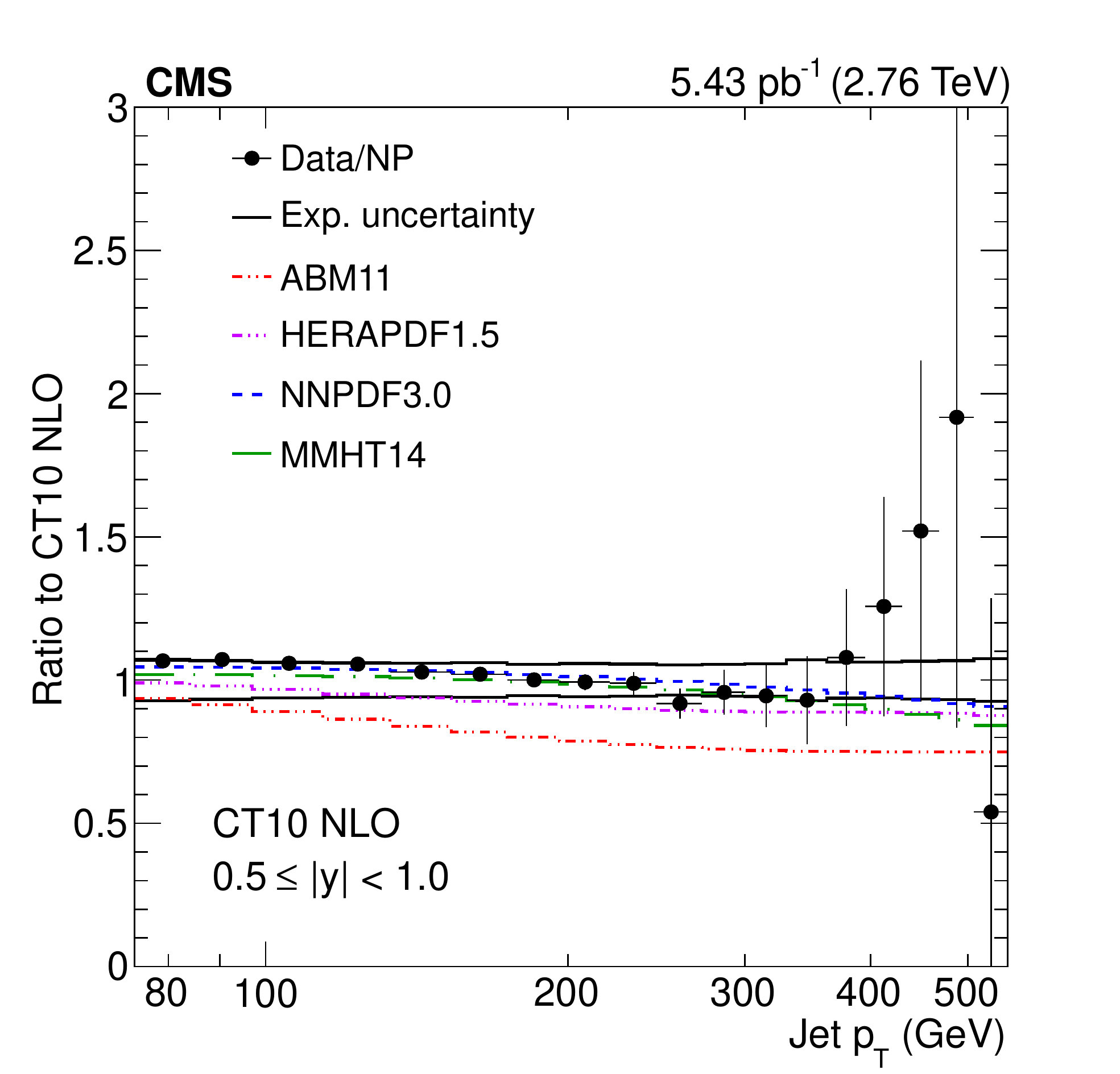} \\
\includegraphics[scale=0.35]{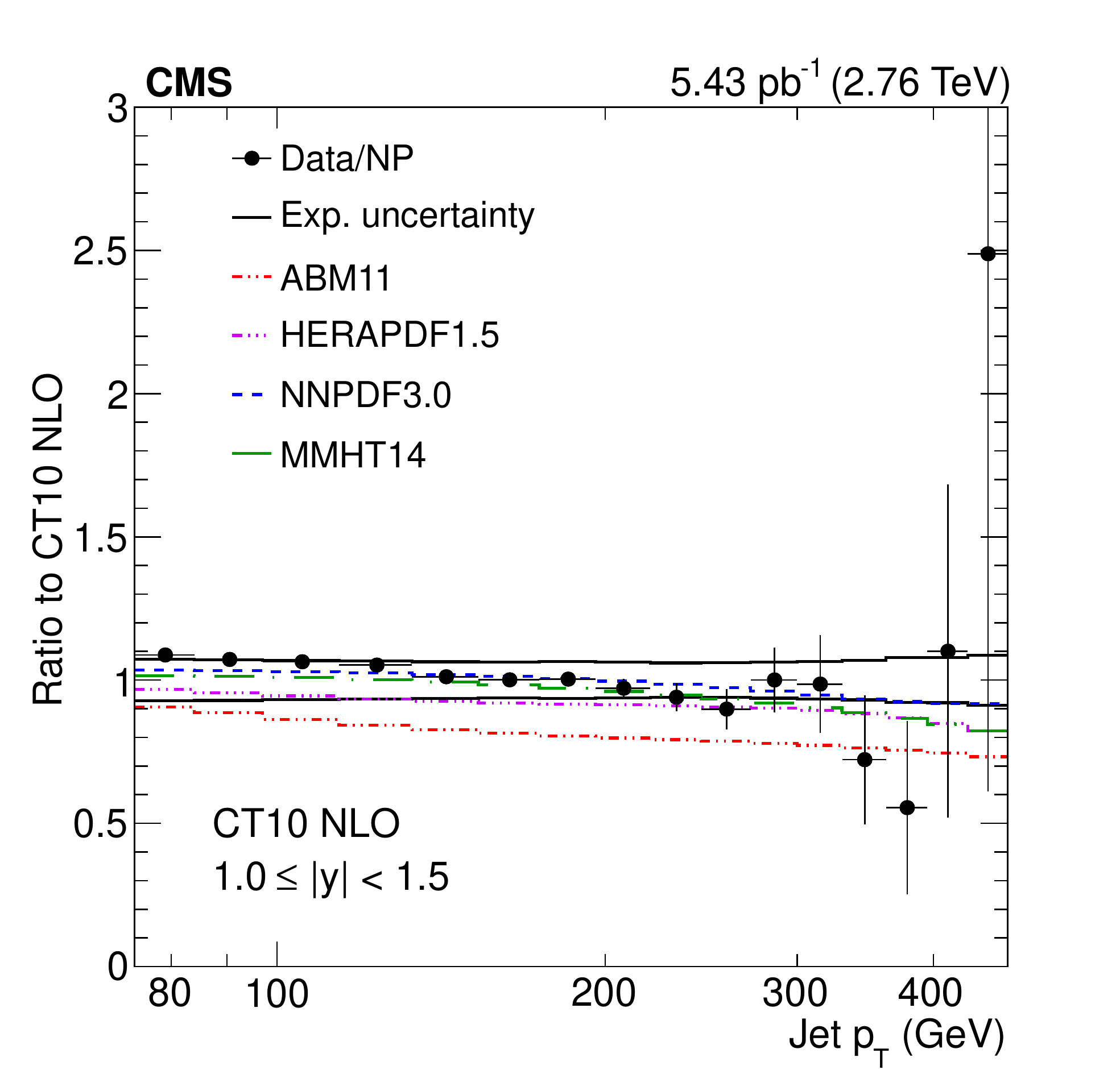}
\includegraphics[scale=0.35]{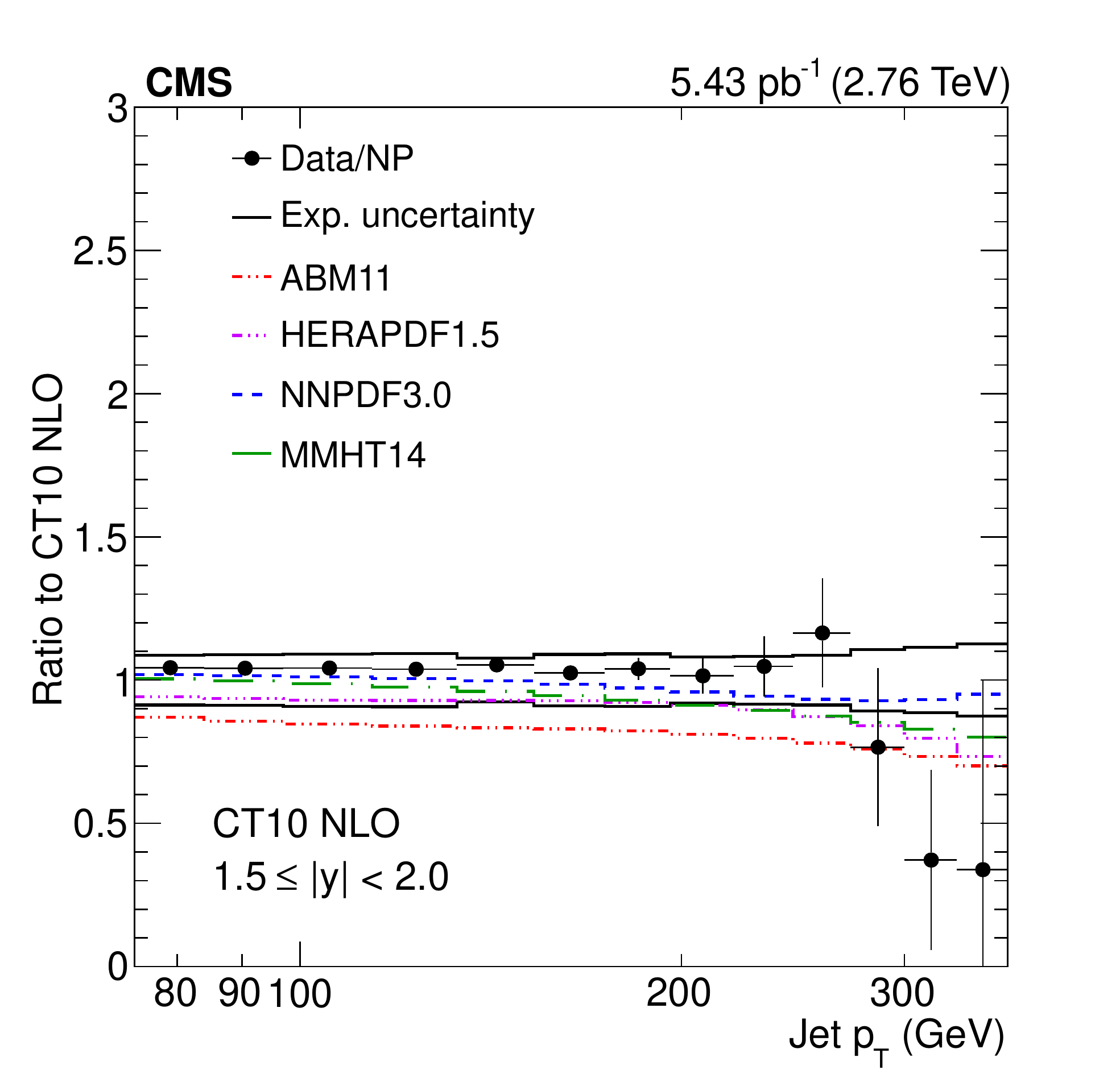} \\
\includegraphics[scale=0.35]{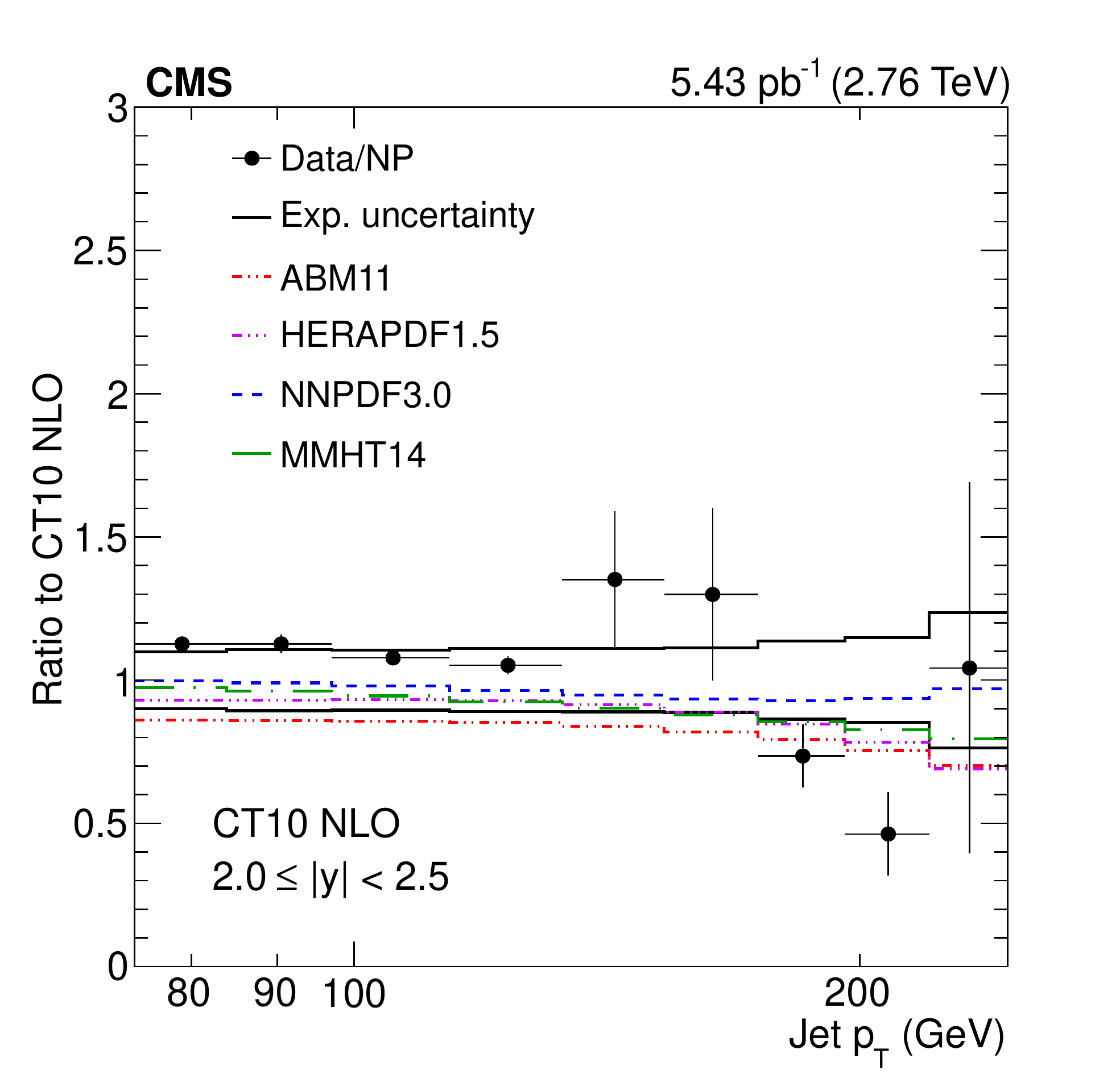}
\includegraphics[scale=0.35]{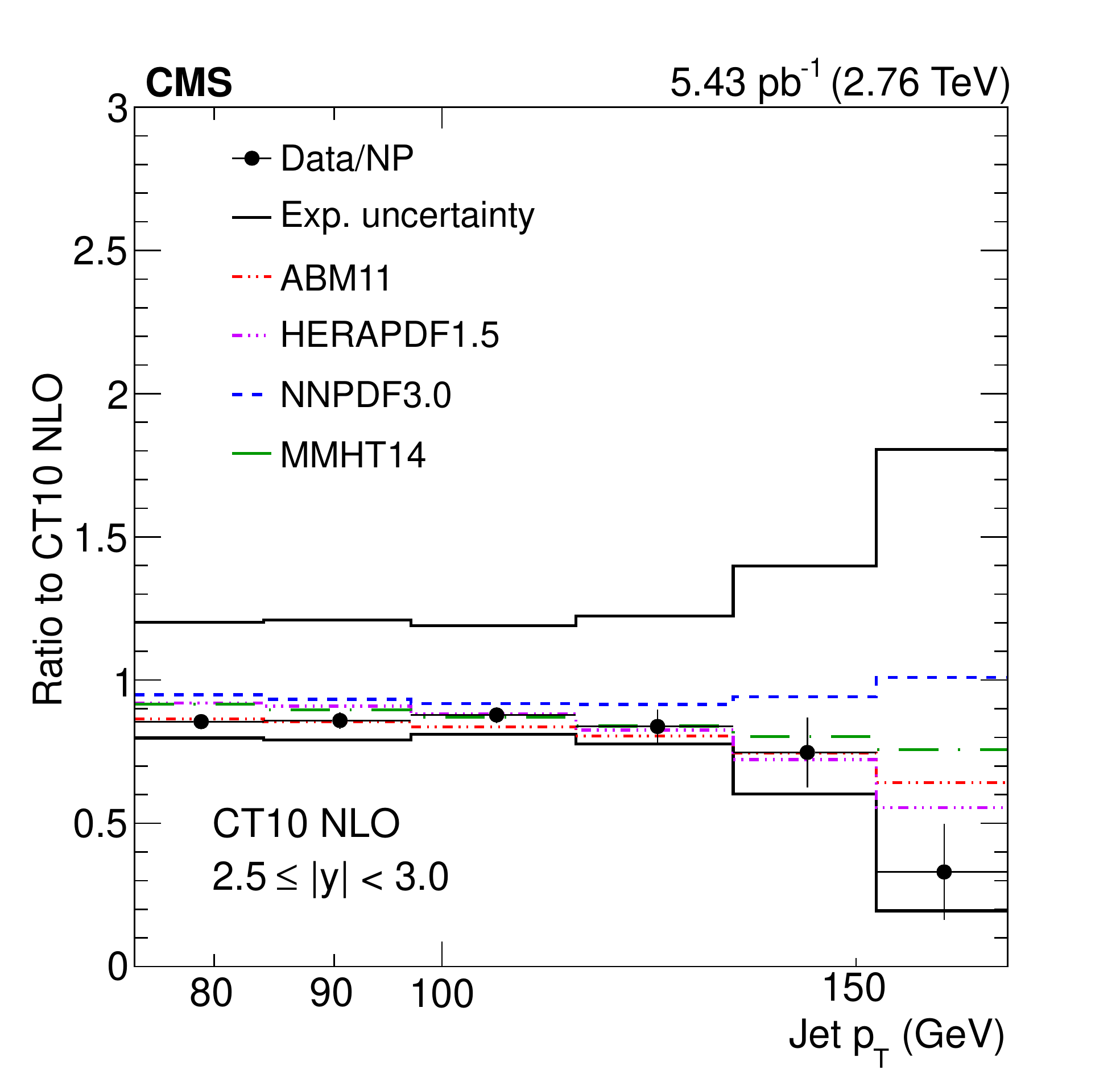}
\caption{The same data shown in Fig. 2 are presented showing comparisons to
the NLO QCD predictions using a variety of PDFs, which are denoted by different
line styles. The uncertainties corresponding to the QCD predictions are not shown.
For simplicity, the NP corrections needed for the various QCD predictions have been
applied to the data in this figure.
}
\label{datathyct10pdfs}
\end{figure*}

The measured inclusive jet cross section and the theoretical  predictions
are compared in Figs.~\ref{combct10}--\ref{datathyct10pdfs}. In Fig. \ref{combct10}, the
double-differential cross section is plotted as a function of jet \pt
and $\abs{y}$. The theoretical prediction obtained with
the CT10 PDF set is shown as well. A more detailed comparison for all $\abs{y}$ bins is
presented in Fig.~\ref{datathyct10}, where the ratios of data to theory using the CT10 PDF set
are shown. Within the uncertainties, the data are well described by NLO QCD in the
full kinematic range explored. In Fig.~\ref{datathyct10pdfs}, the data, with NP corrections,
are compared in a similar manner to the predictions from other PDF sets,
normalized to the CT10 prediction. In general, all predictions describe the
data well. Within experimental and theoretical (not shown) uncertainties,
only the comparison to the prediction from the ABM11 PDF set exhibits slight differences
between the data and theory, an effect that has been observed also in other
measurements, e.g. Ref.~\cite{cmspdf}.

\section{Summary}
A measurement of the double-differential inclusive jet cross section was presented.
The data were collected by the CMS detector in  pp collisions at
 $\sqrt{s} = 2.76\TeV$, with an integrated luminosity of 5.43\pbinv.
The measurement covers the jet kinematic ranges of $74\leq\pt<592\GeV$ and $\abs{y}<3.0$.

A detailed study of the experimental and theoretical uncertainties has been performed.
Contributions to the experimental systematic uncertainty were evaluated
from the jet energy corrections, jet energy resolution, and integrated luminosity.
Jet energy corrections dominate the experimental uncertainty, followed
by smaller contributions from  jet energy resolution and luminosity.
The theoretical uncertainty is dominated by the missing higher-order corrections
that were estimated by varying the renormalization and factorization
scales, and the PDF uncertainty; the contribution of nonperturbative correction
uncertainty is small.

The data are corrected for  detector resolution and efficiencies.
The measured cross sections are compared to NLO QCD predictions
obtained using different PDF sets.
These cross section measurements test and confirm the predictions
of QCD at $\sqrt{s}= 2.76\TeV$
and extend the kinematic range compared to previous studies.

\begin{acknowledgments}

\hyphenation{Bundes-ministerium Forschungs-gemeinschaft Forschungs-zentren} We congratulate our colleagues in the CERN accelerator departments for the excellent performance of the LHC and thank the technical and administrative staffs at CERN and at other CMS institutes for their contributions to the success of the CMS effort. In addition, we gratefully acknowledge the computing centers and personnel of the Worldwide LHC Computing Grid for delivering so effectively the computing infrastructure essential to our analyses. Finally, we acknowledge the enduring support for the construction and operation of the LHC and the CMS detector provided by the following funding agencies: the Austrian Federal Ministry of Science, Research and Economy and the Austrian Science Fund; the Belgian Fonds de la Recherche Scientifique, and Fonds voor Wetenschappelijk Onderzoek; the Brazilian Funding Agencies (CNPq, CAPES, FAPERJ, and FAPESP); the Bulgarian Ministry of Education and Science; CERN; the Chinese Academy of Sciences, Ministry of Science and Technology, and National Natural Science Foundation of China; the Colombian Funding Agency (COLCIENCIAS); the Croatian Ministry of Science, Education and Sport, and the Croatian Science Foundation; the Research Promotion Foundation, Cyprus; the Ministry of Education and Research, Estonian Research Council via IUT23-4 and IUT23-6 and European Regional Development Fund, Estonia; the Academy of Finland, Finnish Ministry of Education and Culture, and Helsinki Institute of Physics; the Institut National de Physique Nucl\'eaire et de Physique des Particules~/~CNRS, and Commissariat \`a l'\'Energie Atomique et aux \'Energies Alternatives~/~CEA, France; the Bundesministerium f\"ur Bildung und Forschung, Deutsche Forschungsgemeinschaft, and Helmholtz-Gemeinschaft Deutscher Forschungszentren, Germany; the General Secretariat for Research and Technology, Greece; the National Scientific Research Foundation, and National Innovation Office, Hungary; the Department of Atomic Energy and the Department of Science and Technology, India; the Institute for Studies in Theoretical Physics and Mathematics, Iran; the Science Foundation, Ireland; the Istituto Nazionale di Fisica Nucleare, Italy; the Ministry of Science, ICT and Future Planning, and National Research Foundation (NRF), Republic of Korea; the Lithuanian Academy of Sciences; the Ministry of Education, and University of Malaya (Malaysia); the Mexican Funding Agencies (CINVESTAV, CONACYT, SEP, and UASLP-FAI); the Ministry of Business, Innovation and Employment, New Zealand; the Pakistan Atomic Energy Commission; the Ministry of Science and Higher Education and the National Science Centre, Poland; the Funda\c{c}\~ao para a Ci\^encia e a Tecnologia, Portugal; JINR, Dubna; the Ministry of Education and Science of the Russian Federation, the Federal Agency of Atomic Energy of the Russian Federation, Russian Academy of Sciences, and the Russian Foundation for Basic Research; the Ministry of Education, Science and Technological Development of Serbia; the Secretar\'{\i}a de Estado de Investigaci\'on, Desarrollo e Innovaci\'on and Programa Consolider-Ingenio 2010, Spain; the Swiss Funding Agencies (ETH Board, ETH Zurich, PSI, SNF, UniZH, Canton Zurich, and SER); the Ministry of Science and Technology, Taipei; the Thailand Center of Excellence in Physics, the Institute for the Promotion of Teaching Science and Technology of Thailand, Special Task Force for Activating Research and the National Science and Technology Development Agency of Thailand; the Scientific and Technical Research Council of Turkey, and Turkish Atomic Energy Authority; the National Academy of Sciences of Ukraine, and State Fund for Fundamental Researches, Ukraine; the Science and Technology Facilities Council, UK; the US Department of Energy, and the US National Science Foundation.

Individuals have received support from the Marie-Curie program and the European Research Council and EPLANET (European Union); the Leventis Foundation; the A. P. Sloan Foundation; the Alexander von Humboldt Foundation; the Belgian Federal Science Policy Office; the Fonds pour la Formation \`a la Recherche dans l'Industrie et dans l'Agriculture (FRIA-Belgium); the Agentschap voor Innovatie door Wetenschap en Technologie (IWT-Belgium); the Ministry of Education, Youth and Sports (MEYS) of the Czech Republic; the Council of Science and Industrial Research, India; the HOMING PLUS program of the Foundation for Polish Science, cofinanced from European Union, Regional Development Fund; the OPUS program of the National Science Center (Poland); the Compagnia di San Paolo (Torino); MIUR project 20108T4XTM (Italy); the Thalis and Aristeia programs cofinanced by EU-ESF and the Greek NSRF; the National Priorities Research Program by Qatar National Research Fund; the Rachadapisek Sompot Fund for Postdoctoral Fellowship, Chulalongkorn University (Thailand); the Chulalongkorn Academic into Its 2nd Century Project Advancement Project (Thailand); and the Welch Foundation, contract C-1845.

\end{acknowledgments}

\clearpage

\bibliography{auto_generated}

\cleardoublepage \appendix\section{The CMS Collaboration \label{app:collab}}\begin{sloppypar}\hyphenpenalty=5000\widowpenalty=500\clubpenalty=5000\textbf{Yerevan Physics Institute,  Yerevan,  Armenia}\\*[0pt]
V.~Khachatryan, A.M.~Sirunyan, A.~Tumasyan
\vskip\cmsinstskip
\textbf{Institut f\"{u}r Hochenergiephysik der OeAW,  Wien,  Austria}\\*[0pt]
W.~Adam, E.~Asilar, T.~Bergauer, J.~Brandstetter, E.~Brondolin, M.~Dragicevic, J.~Er\"{o}, M.~Flechl, M.~Friedl, R.~Fr\"{u}hwirth\cmsAuthorMark{1}, V.M.~Ghete, C.~Hartl, N.~H\"{o}rmann, J.~Hrubec, M.~Jeitler\cmsAuthorMark{1}, V.~Kn\"{u}nz, A.~K\"{o}nig, M.~Krammer\cmsAuthorMark{1}, I.~Kr\"{a}tschmer, D.~Liko, T.~Matsushita, I.~Mikulec, D.~Rabady\cmsAuthorMark{2}, B.~Rahbaran, H.~Rohringer, J.~Schieck\cmsAuthorMark{1}, R.~Sch\"{o}fbeck, J.~Strauss, W.~Treberer-Treberspurg, W.~Waltenberger, C.-E.~Wulz\cmsAuthorMark{1}
% ...
\vskip\cmsinstskip
% ...
1:~~Also at Vienna University of Technology, Vienna, Austria\\
2:~~Also at CERN, European Organization for Nuclear Research, Geneva, Switzerland\\
% ...

\end{sloppypar}
\end{document}